\documentclass[aps,prl,twocolumn,superscriptaddress,amsmath,amssymb]{revtex4-1}

\usepackage{xcolor}
\usepackage{graphicx}
\usepackage{bm}
\usepackage{lineno}
\usepackage{multirow}
\usepackage{footnote}
\usepackage{url,hyperref}
\usepackage{soul}
\usepackage{amsmath}
\usepackage[utf8]{inputenc}
\usepackage[T1]{fontenc}

\newcommand{\qq}{\mathbf{q}}
\newcommand{\rr}{\mathbf{r}}
\newcommand{\GG}{\mathbf{G}}
\newcommand{\Gz}{G_z}
\newcommand{\qz}{q_z}

\newcommand{\w}{\omega}

\newcommand{\qp}{\mathbf{q}_\parallel}
\newcommand{\Gp}{\mathbf{G}_\parallel}
\newcommand{\Gpp}{\mathbf{G}'_\parallel}

\newcommand{\kp}{\mathbf{k}_\parallel}
\newcommand{\Kp}{\bm{\kappa}_\parallel}
\newcommand{\nn}{\nonumber}
\newcommand{\IN}{\text{in}}
\newcommand{\OFF}{\text{off}}
\newcommand{\lb}[1]{\left( {#1} \right|}
\newcommand{\rb}[1]{\left| {#1} \right)}
\newcommand{\lrbb}[2]{\left(  \left.{#1} \right| {#2} \right)}
\newcommand{\lrb}[2]{\left(  {#1} \left|  {#2} \right. \right)}
\newcommand{\lrbbb}[3]{\left( {#1} \left| {#2} \right| {#3} \right)}

\begin{document}

\title{Proper ab-initio dielectric function of 2D materials and their polarizable thickness}

\author{Lorenzo Sponza}
\affiliation{Laboratoire d'Etude des Microstructures, ONERA-CNRS, UMR104, Universit\'e Paris-Saclay, BP 72, 92322 Ch\^atillon Cedex, France}

\author{Fran\c{c}ois Ducastelle}
\affiliation{Laboratoire d'Etude des Microstructures, ONERA-CNRS, UMR104, Universit\'e Paris-Saclay, BP 72, 92322 Ch\^atillon Cedex, France}

\begin{abstract}
In this paper we derive a formalism allowing us to separate inter-layer contributions to the polarizability of a periodic array of 2D materials from intra-layer ones. To this aim, effective  profile functions are introduced. They constitute a tight-binding-like layer-localized basis involving two lengths, the effective thickness $d$ characteristic of the 2D material and the inter-layer separation $L$. The method permits, within the same formalism, either to compute the single-layer dielectric function from an ab-initio periodic calculation (top-down strategy) or to stack several 2D materials to generate a finite-thickness van der Waals heterostructure (bottom-up strategy).
\end{abstract}

%%% \date{\today}

\maketitle

Most ab initio  codes assume periodic boundary conditions in the 3D space.
This framework is inappropriate to simulate isolated systems, interfaces, defects, amorphous materials and any non-periodic or partially periodic system. Isolated 2D sheets fall in this category, with a crystalline structure periodic in one plane (the $xy$-plane, in the following) and isolated along the vertical direction $z$. Similarly, the simulation of multilayers or van der Waals heterostructures  is often prohibitive because of the number of atoms involved, even when lattice mismatch or misalignment between the constituent layers are neglected.
In this paper we address both problems within the same formalism in a fully ab initio approach.

The first problem is about the calculation of single-layer properties from 3D-periodic ab initio calculations (top-down strategy).
In this framework, fictitious Coulomb interactions arise between the replicas of the system. The brute force approach consists in creating large simulation cells (supercell: SC) enclosing the planar unitary cell plus an amount of vacuum large enough to separate the periodic replicas of the system. A vacuum of $\sim$10~\AA{ } is often required to converge regular ground-state simulations.
Though, when it comes to computing the polarizability $\chi$, for instance in the random phase approximation (RPA), local dipoles are created. These charge inhomogeneities worsen the convergence problem because the artifact interactions are unscreened and hence long-ranged. Furthermore $\chi$ is normalized with respect to the volume of the simulation cell, and hence depends on the amount of vacuum which is arbitrary~\cite{Cudazzo2011,Sponza2016,Tian2020}. In our opinion, this renormalization problem has been too much overlooked in the past; mainly because it is absent in ground-state calculations. As a result, in the SC scheme, excited state calculations seldom converge with the amount of vacuum.

However, state-of-the-art calculations are nowadays done within the Coulomb truncation (CT) scheme \cite{ismail-beigi_prb2006,rozzi_prb2006,huser_prb2013,cudazzo_prl2016}. It consists in letting the Coulomb interaction vanish at a distance $L/2$ from the sheet, where $L$ is the height of the simulation cell. This scheme kills by construction the spurious interactions and permits to reduce $L$ to values comparable to those needed in SC ground-state simulations. Though, we want to stress that $L$ is still an arbitrary quantity even in the CT scheme.

The second problem consists in stacking different 2D layers to get finite-thickness homo- or hetero-structures (bottom-up strategy). 
A solution to the two problems must rely on the correct description of the inter-layer interaction and also on the explicit account for the (effective) thickness of the layers.

In this Letter we present a general formalism allowing us to separate analytically the intra-layer from inter-layer contributions to the polarizability $X$ of a periodic array of layers. This is done by defining profile functions while computing the polarizability of a single layer. In this way, the volume-normalization and the fictitious-interaction problems are treated on the same footing, completing and generalizing recent developments~\cite{nazarov_prb2014,nazarov_njp2015,latini_prb2015, thygesen_2D2017,Meckbach2018a,Tian2020,Gjerding2020} concerning the top-down strategy. Then, we consider the general case of a finite-thickness heterostructure and we use the same general framework to implement a bottom-up strategy similar to the Quantum Electrostatic Heterostructure (QEH) model recently put forward~\cite{andersen_nl2015,thygesen_2D2017,Gjerding2020,Cavalcante2018a}.
With the intent of removing any arbitrariness from the calculation, we also provide a recipe to compute the profile functions ab initio and use them to calculate the macroscopic dielectric function of a single layer. This development allows us to demonstrate that for the dielectric function to be meaningful, it is mandatory to take into account the finite thickness of the film.

\paragraph{I: Single-layer polarizability}
We start from the case of an isolated single layer without any periodicity along  the $z$ axis.
We introduce a mixed space representation $\rb{\kp,z}$ for which a generic quantity depends on the in-plane momentum ($\kp=\qp+\Gp$, sum of the crystal momentum $\qp$ and of a reciprocal lattice vector $\Gp$~
\footnote{Note that in this notation, the continuous part never changes. So if $\kp=\qp+\Gp$, then $\kp'=\qp+\Gp'$ and analogously $k_z' = q_z+G_z'$.}), and on the real-space coordinate $z$.
We proceed assuming that, we can separate in-plane ($\varphi$) from out-of-plane ($\xi$) components of the Bloch wavefunction:  $\phi_{i \Kp}(\rr)= \varphi_{i \Kp}(\bm{\rho})\,\xi(z)$, where $\bm{\rho}$ is the in-plane component of $\rr$.
The irreducible polarizability $\chi^0$ is calculated as usual as a sum over all independent-particle transitions, the oscillator strengths of which involve matrix elements of the form $\int \phi^*_{i \Kp}(\rr) e^{i (\qp+\Gp) \cdot \bm{\rho} } \phi_{j \Kp+\qp}(\rr)\, d \bm{\rho}$.
This makes appear the normalized effective profile functions $\theta^0=\xi^2$ which are positive, real and even functions of $z$ and localized around $z=0$ (i.e. they vanish for $|z|>d/2$ for a given distance $d$). This leads to the expression for the irreducible polarizability of the single layer:
\begin{equation}
\chi^0_{\Gp\Gp'}(\qp,\w,z,z') = \theta^0(z) \chi^0_{2D \Gp\Gp'}(\qp,\w) \theta^0(z') \,.
\label{eq:chi02d_mixed-rep_1}
\end{equation} 
The functions $\theta^0$ play here the role of effective polarizable profiles and they can be associated to the thickness of the isolated layer through the characteristic length $d$. This thickness is defined more precisely below.
It turns out that the numerical extraction of the profile function shows a mild dependence on the in-plane momentum, and in the following
we include such a dependence. Then, $\theta^0_{\kp}(z)$ is characterized by an effective thickness $d_{\kp}$.
Since our factorization hypothesis is done at the wave-function level, we assume no $\w$-dependence of $\theta^0$.

The random phase approximation (RPA) to the polarizability of the slab $\chi$ satisfies a Dyson equation of the form $\chi = \chi^0 + \chi^0 v \chi$. As a consequence of the $\theta^0$ functions appearing in~\eqref{eq:chi02d_mixed-rep_1}, one can show that $\chi = \theta^0  \chi_{2D}  \theta^0$ where the extension in $(z,z')$ is carried only by the $\theta^0$ functions, while the polarization is embodied in  $\chi_{2D\,\Gp\Gp'}(\qp,\w)$ which solves the 2D Dyson equation:
\begin{equation}
\chi_{2D} = \chi^0_{2D} + \chi^0_{2D}\mathcal{V}^{\IN}\chi_{2D}\,.
\label{eq:single-layer_dyson}
\end{equation}
Here $\mathcal{V}^{\IN}$ is the layer-projected Coulomb interaction;
\begin{equation}
\mathcal{V}^{\IN}(\kp) = \frac{2\pi}{|\kp|} \int \theta^0_{\kp}(z) e^{-|\kp| \, |z-z'|} \theta^0_{\kp'}(z') \,dz\, dz'\,. 
\label{eq:vin}
\end{equation}
For further details on the entire derivation, see appendix A of the Supplemental Material.\\

\paragraph{II: Periodic array of layers}
In order to study the polarizability $X$ of an array of layers with period $L$ along $z$, we introduce a layer-projected representation $\rb{\kp,n}$ where $n$ is the index of the layer.
We define $X_{\Gp,\Gp',nm}(\qp,\w) = \lb{\kp,n}X(\w)\rb{\kp',m}$ and the projection from $\lrb{\kp,z}{\kp',n}    =  \sqrt{d_{\kp}}\theta^n_{\kp}(z) \delta_{\Gp \Gp'}$, where $\theta^n_{\kp}(z):=\theta^{0}_{\kp}(z-nL)$ is the profile function localized on the $n$th layer. 
Details on the representations can be found in  Appendix B of the Supplemental Material. Since $\chi^0$ does not imply any long-range term, the overlap of successive layers can be neglected if we further assume $L>d_{\kp} \, \forall \,\kp$.
So $X^0$ becomes the sum over the periodically repeated $\chi^0$s (equation~\eqref{eq:chi02d_mixed-rep_1}):
\begin{eqnarray}
\lefteqn{ X^0_{\Gp\Gp'}(\qp,\w,z,z') =} \nn \\ 
&\qquad =&\sum_{n} \theta^n_{\kp}(z) \chi^0_{2D\,\Gp\Gp'}(\qp,\w)\theta^n_{\kp'}(z')   \label{eq:sum-over-single-layers}  \\
&\qquad =&\sqrt{d_{\kp}d_{\kp'}} \sum_{n,m}  \theta^n_{\kp}(z)X^0_{\Gp\Gp',nm}(\qp,\w) \theta^m_{\kp'}(z') \nn \;,
\end{eqnarray}
depending on the representation employed (mixed-space above, or layer-projected below).
By equating these two expressions one gets $\sqrt{d_{\kp}d_{\kp'}} X^0_{\kp\kp'\,nm} = \delta_{nm} \chi^0_{2D\,\kp\kp'}$.
Using this equality 
and by defining $\hat{\chi}^{(0)}_{\kp\kp'}:=\sqrt{d_{\kp}d_{\kp'}} X^{(0)}_{\kp\kp'}$ and $ \hat{v}(\kp)=v(\kp)/d_{\kp}$, one arrives to the layer-projected Dyson equation: 
\begin{equation}
\hat{\chi}_{nm} = \hat{\chi}^0_{nm} + \sum_{p,s}\hat{\chi}^0_{np} \hat{v}_{ps} \hat{\chi}_{sm}\, .
\label{eq:dyson_layer-projected}
\end{equation}
We notice that $\hat{v}_{nn}(\kp) = \lb{\kp,n}\hat{v}\rb{\kp,n}\equiv\mathcal{V}^{\IN}(\kp)$
as in~\eqref{eq:vin}. Using the definitions introduced above, the $n$th layer term reads $\hat{\chi}_{nn} = \chi^0_{2D} + \chi^0_{2D}\mathcal{V}^{\IN}\chi^0_{2D}+O(2)$. We conclude that $\hat{\chi}_{nn}\equiv\chi_{2D}$ which is  layer-independent and solves the single-layer equation~\eqref{eq:single-layer_dyson}. Note that all matrices are written in  $\Gp$ space,  so that in-plane local fields are fully taken into account.

Let us now focus on the off-diagonal elements of~\eqref{eq:dyson_layer-projected}.
We start splitting the Coulomb interaction $\hat{v}_{nm} = \mathcal{V}^{\IN}_{nm} + \mathcal{V}^{\OFF}_{nm}$ 
with the definitions $\mathcal{V}^{\IN}_{nm} = \delta_{nm}\mathcal{V}^{\IN}$, and $\mathcal{V}^{\OFF}_{nm}=\hat{v}_{nm}$ if $n\neq m$ and 0 otherwise.
Equation~\eqref{eq:dyson_layer-projected} can now be split into two  equations:
\begin{eqnarray}
\chi_{2D,nm} &=& \delta_{nm}  \left[ \chi_{2D}^0 \left(1-\mathcal{V}^{\IN} \,  \chi_{2D}^0\right)^{-1}\right] = \delta_{nm}\chi_{2D}\\
\hat{\chi}_{nm}  &=& \delta_{nm}\chi_{2D} + \sum_{q}\chi_{2D}  \mathcal{V}^{\OFF}_{nq} \, \hat{\chi}_{qm}\,. \label{eq:dyson_off-diagonal}
\end{eqnarray}
The latter equation is the main result: It expresses the total polarizability of the array ($\hat{\chi}$) in terms of single-layer polarizabilities ($\chi_{2D}$) interacting through an effective inter-layer Coulomb interaction ($\mathcal{V}^{\OFF}$). This result permits to develop the bottom-up and the top-down strategies presented previously. The advantage of our approach is to combine them within a general and concise formalism based in particular on the profile functions $\theta^0$ and their associate polarizable thickness. They are crucial quantities for a correct calculation of the dielectric function. We will show how to compute them from first principles, hence removing any arbitrariness in ab initio calculations of single layers.

With the intent of connecting our derivation to standard ab initio output, usually expressed in the $\rb{\kp,k_z}$ representation, we  rewrite~\eqref{eq:dyson_off-diagonal} in reciprocal space. 
This is done in detail in  Appendix C of the Supplemental Material; below we report just the main results for $q_z=0$. $\w$ and $\qq=\qp$ variables are dropped for notational convenience:
\begin{eqnarray}
X_{\GG\GG'} &=& L^{-1} \vartheta_{\kp}(G_z) \hat{\chi}_{\Gp\Gp'} \vartheta_{\kp'}(G_z')  \label{eq:solution_tildechi}\\
\chi_{2D,\Gp\Gp'} & = & \hat{\chi}_{\Gp\Gp'} - \chi_{2D,\Gp\Gp''}\mathcal{V}^{\OFF}(\kp'')\hat{\chi}_{\Gp''\Gp'}  \label{eq:solution_dyson} \\ 
 \mathcal{V}^{\OFF}(\kp'') &=&   2 \mathcal{V}^{\IN}(\kp'') \left[ e^{|\kp''| L} - 1 \right]^{-1}\,. \label{eq:solution_voff_1}
\end{eqnarray}
In~\eqref{eq:solution_tildechi}, $\vartheta$ is the Fourier transform along $z$ of the profile function $\theta^0$. A sum over $\Gp''$ is understood in~\eqref{eq:solution_dyson}.

Here we present the practical implementation of the top-down method. We keep assuming $q_z=0$. The first step is to extract the profile function $\theta^0$ from the first-principle periodic calculation of $X$. By sampling the column $X_{\GG,0}(\qp,\w=0)$ along $\Gz$ at fixed $\Gp$, one can extract $\theta^0_{\kp}$ from first principles:
\begin{equation}
\vartheta_{\kp}(\Gz) = \frac{X_{\GG\, 0}(\qp,0)}{X_{\Gp\,0}(\qp,0) }  \; ,
\label{eq:profiles}
\end{equation}
from which we obtain  $\theta_{\kp}^0(z) $,  the Fourier transform of $\vartheta_{\kp}(k_z)$ which, for large $L$, can be calculated through a discrete sum $\theta_{\kp}^0(z)  \simeq \frac{1}{L}\sum_{G_z} e^{iG_{z}z} \vartheta_{\kp}(G_z)$.
Note that this direct relation between the profile function and the polarizability confirms the close relation identified already by Tian and coworkers~\cite{Tian2020}. However, our ab initio method prevents the use of an arbitrary thickness parameters to be fixed from fitting procedures~\cite{Meckbach2018a} or from physical considerations~\cite{latini_prb2015}.

Once the $\vartheta$s extracted, one can invert equation~\eqref{eq:solution_tildechi} at $\w=0$ thus getting $\hat{\chi}$ from first principles.
Next, we calculate the inter-layer Coulomb interaction $\mathcal{V}^{\OFF}$ as defined in~\eqref{eq:solution_voff_1}. 
This can be done numerically (NUM) relying on~\eqref{eq:vin}, or can be approximated.
Within the \emph{perfect 2D} approximation (2DA), $\theta^0_{\kp}(z) = \delta(z)$.
In the \emph{sharp slab} approximation (SSA) $\theta^0_{\kp}(z) = d^{-1} \left[ H\left(z+d/2\right) - H\left(z-d/2\right)\right]$ where $H(x)$ is the Heaviside step function and $d$ is an effective momentum-independent thickness.
Depending on the the approximation chosen, 
\begin{equation}
\mathcal{V}^{\IN}(\kp) = \left\{ 
\begin{array}{ll}
2\pi/|\kp| & \text{2DA}  \\
\frac{4\pi}{|\kp|^3d^2} \left(|\kp|d - 1 + e^{-|\kp|d}\right) & \text{SSA} \\
\text{computed as in~\eqref{eq:vin}} & \text{NUM} 
\end{array}
\right.
\label{eq:vin_approx}
\nn
\end{equation}
Other approximations are possible (for instance the 2D Ohno potential as in~\cite{Meckbach2018a}).
Putting it into~\eqref{eq:solution_voff_1} and solving~\eqref{eq:solution_dyson}, one finally gets $\chi_{2D}$.
We stress that, if $X$ has been computed in the CT scheme, then $\mathcal{V}^{\OFF}\equiv 0$ and $\hat{\chi}\equiv\chi_{2D}$.

In the bottom-up method we calculate the global polarizability of an heterostructure composed of $N$ layers of different 2D materials. 
Equation~\eqref{eq:dyson_off-diagonal} can be easily generalized to this case and becomes:
\begin{equation}
\hat{\chi}_{nm}  = \delta_{nm}\chi_{2D,n} + \sum_{p=1}^N\chi_{2D,n}  \mathcal{V}^{\OFF}_{np} \, \hat{\chi}_{pm} \,, \label{eq:heterostructure_1} 
\end{equation}
with
\begin{equation}
\mathcal{V}^{\OFF}_{nm}(\kp) = \int \int \theta^0_{\kp}(z-z_n) v(\kp,z,z')\theta^0_{\kp'}(z'-z_m) dz dz' \,, \label{eq:heterostructure_2} 
\end{equation}
where the differences with ~\eqref{eq:dyson_off-diagonal} are that (i) the sum is limited to $N$ layers, (ii) each layer has its own $\chi_{2D,n}$, and (iii) the generic layer $n$ centered at $z_n$ is not necessary in a periodic array.
%\LS{\st{if layer $n$ and layer $m$ are done of the same material, then $\chi_{2D,n}=\chi_{2D,m}$.}}
The bottom-up equation~\eqref{eq:heterostructure_1} can be applied to particular cases. For instance, in a multilayer made of the same 2D material $\chi_{2D,n} = \chi_{2D} \, \forall n$.  Also the bulk can be reconstructed by adding in the latter case $N=\infty$ and $z_n = nL$, which actually boils down in solving the pristine equation~\eqref{eq:dyson_off-diagonal}.

\paragraph{III: The dielectric function}

The inverse microscopic dielectric function of a single layer $\epsilon^{-1}$ relates the total potential $U^\text{tot}$ to an external potential $U^\text{ext}$ according to the definition $ U^{\text{tot}} =   \epsilon^{-1} U^{\text{ext}} $. The macroscopic average of these fields is obtained through their projection on the single layer. 
To do so, we project them on the $\rb{\kp,n=0}$ representation (see Supp Mat Appendix B), getting $ U^j_M(\qp,\w) = \sqrt{d_{\kp}} \int \theta^0_{\kp}(z) U^j_{0}(\qp,\w,z) dz  $, for $j=(\text{ext})$ or $(\text{tot})$.
Next, we define the macroscopic dielectric function as the ratio $\epsilon_M= U^{\text{ext}}_M /U^\text{tot}_M $.
If we assume from the beginning that $U^{\text{ext}}$ is macroscopic (i.e. it vanishes where $\GG\neq 0$), 
we make use of~\eqref{eq:chi02d_mixed-rep_1} and of $\epsilon^{-1} = 1+v\chi$, we obtain:
\begin{equation}
\epsilon_M(\qp,\w) = 1/\left[ 1 + \mathcal{V}^\IN(\qp)\chi_{2D\,00}(\qp,\w) \right].
\label{eq:epsilon_M}
\end{equation}
As before, $\mathcal{V}^{\IN}$ can be computed using the profile functions or using some approximations (e.g. 2DA or SSA).
We stop here to stress an important point. The Coulomb term $\mathcal{V}^{\IN}$ appearing in~\eqref{eq:epsilon_M} embodies the interaction between the electrons of the system and an external charge confined in the slab. This is a proper definition of the average internal dielectric constant, adopted also in~\cite{nazarov_prb2014,nazarov_njp2015,latini_prb2015}. This differs from that used by Qiu \emph{et al.}~\cite{Qiu2016}, who defined it as the ratio between the screened and the bare interaction between two charges. When describing energy loss spectra or optical responses, the pertinent macroscopic dielectric function couples the genuine response of the $2D$ system $\chi_{2D\,00}(\qp,\w)$ to an external field. Such a coupling is always  $4\pi/|\qp|^2$ and not $\mathcal{V}^\IN(\qp)$ as above. In other terms, the correct Coulomb potential a priori is not the same in~\eqref{eq:single-layer_dyson} and in \eqref{eq:epsilon_M}. This is also discussed by Nazarov in the case of electron loss spectroscopy~\cite{nazarov_njp2015}. Actually many standard 3D calculations provide values of $\tilde{\epsilon}_M=1/(1+vX)$, where $v$ is either $4\pi/|\qp|^2$ (in the SC scheme) or its truncated version $4\pi \left(1-e^{-|\qp|L/2}\right)/|\qp|^2$ (CT). So, instead of solving the 2D Dyson equation~\eqref{eq:solution_dyson}, it is then possible to extract directly the macroscopic polarisability within the CT scheme using the fact that by definition $\vartheta(0)=1$, which implies that $LX_{00}(\qp,\w) =\hat{\chi}_{00}(\qp,\w)$ :
\begin{equation}
\chi_{2D,00}(\qp,\w)  = \frac{L |\qp|^2 }{4\pi \left(1-e^{-|\qp|L/2}\right)} \left[ \frac{1}{\tilde{\epsilon}_M(\qp,\w)} -1\right]\,,
\label{eq:pp_cutoff}
\end{equation}
which can be finally plugged into~\eqref{eq:epsilon_M}. No additional approximation has been done with respect to the full approach presented so far. Instead, some approximations have to be done when deriving a similar post-processing treatment for SC data. 

Taking the macroscopic limit of equation~\eqref{eq:solution_dyson}, which becomes hence a scalar equation relating $\chi_{2D,00}$ to $\hat{\chi}_{00}$, we adopt the 2DA for $\mathcal{V}^{\OFF}$, and then:
\begin{equation}
\frac{1}{\chi_{2D,00}(\qp,\w)} \approx \frac{4\pi}{|\qp|^2}\left[ \frac{|\qp|}{e^{|\qp|L} - 1} + \frac{\tilde{\epsilon}_M(\qp,\w)}{L\left(1-\tilde{\epsilon}_M(\qp,\w)\right)}\right].
\label{eq:pp_supercell}
\end{equation}
An equation similar to~\eqref{eq:solution_dyson}, has been derived by Nazarov~\cite{nazarov_prb2014, nazarov_njp2015} under the 2DA.
He then derived the same post-processing equation~\eqref{eq:pp_supercell} and  plugged it in the 2DA version of~\eqref{eq:epsilon_M}. Our method generalizes Nazarov's work, making it applicable to any ab initio scheme (SC or CT) and extends it beyond the 2DA.

\paragraph{IV: Applications}

\begin{figure}
\centering
\includegraphics[height=9.0cm]{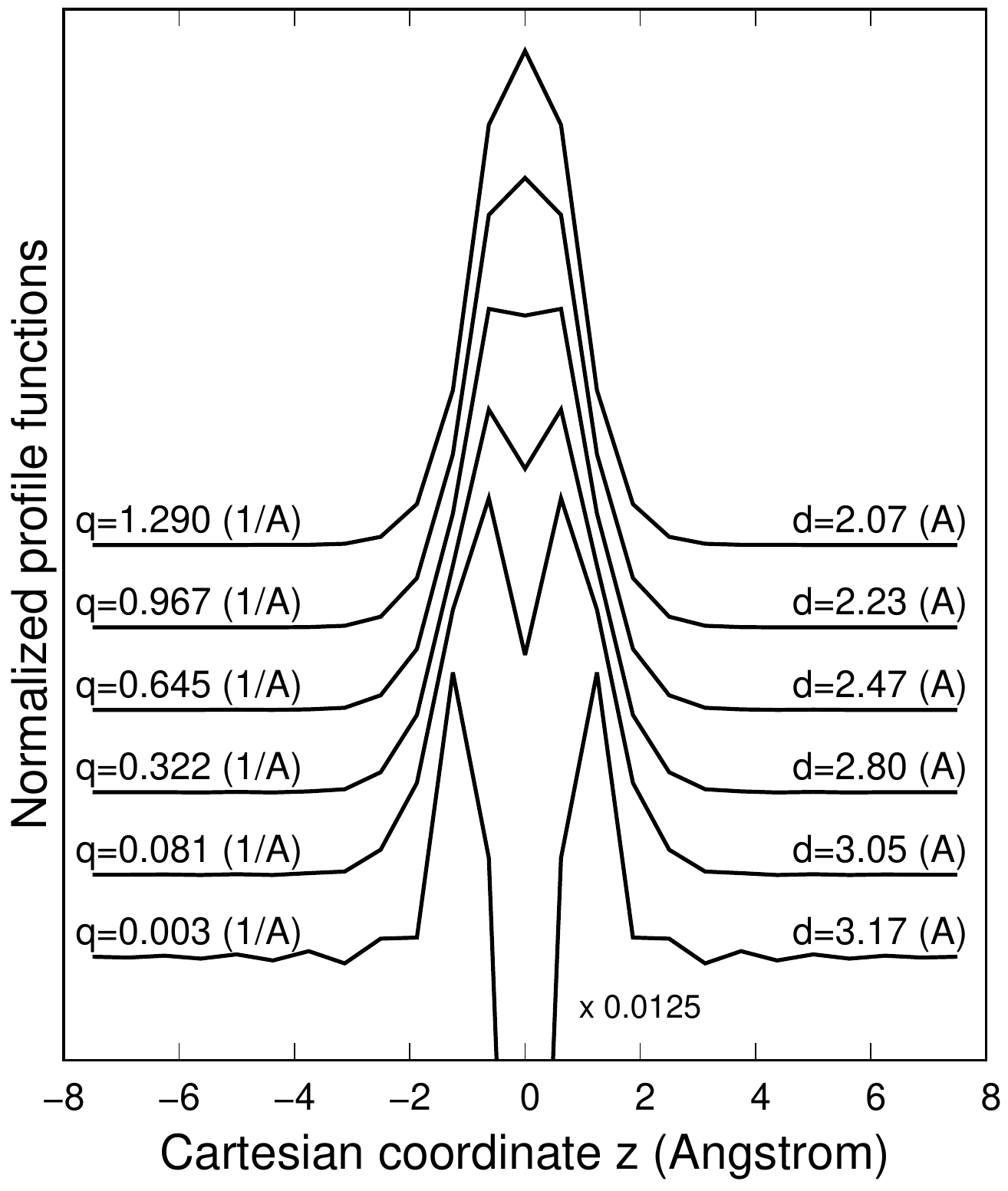}
\caption{Profile function at selected parallel momenta for hBN single layer.
A vertical shift is applied to ease the reading.}
\label{fig:fig1}
\end{figure}

We present here results obtained by applying the top-down strategy to hexagonal boron nitride (hBN).
All calculations have been done with ABINIT \cite{abinit} (wavefunctions), DP \cite{dp} (the $X^0$) and an in-home code implementing our technique. The computational details are presented in Appendix E.

The profile functions $\theta_{\kp}(z)$ extracted as in equations~\eqref{eq:profiles} are reported in Figure~\ref{fig:fig1} for some in-plane momenta (on the right of each profile).
Oscillations come from the numerical evaluation of the Fourier transform, which is the reason for the sharp spike at very small $\qp$. 
We verified however that our results (and in particular $\mathcal{V}^{\IN}$) are stable even in presence of these numerical issues. 
Notice at low $q$ the dip reflecting the $\pi_z$-like electronic density.
Just for comparison, we report also the full with at half maximum of each profile as an estimate of the effective thickness $d_{\kp}$.
We see that this is a bit lower than the inter-layer distance in bulk hBN and shrinks for larger $\kp$.

\begin{figure}
\centering
\includegraphics[width=0.46\textwidth]{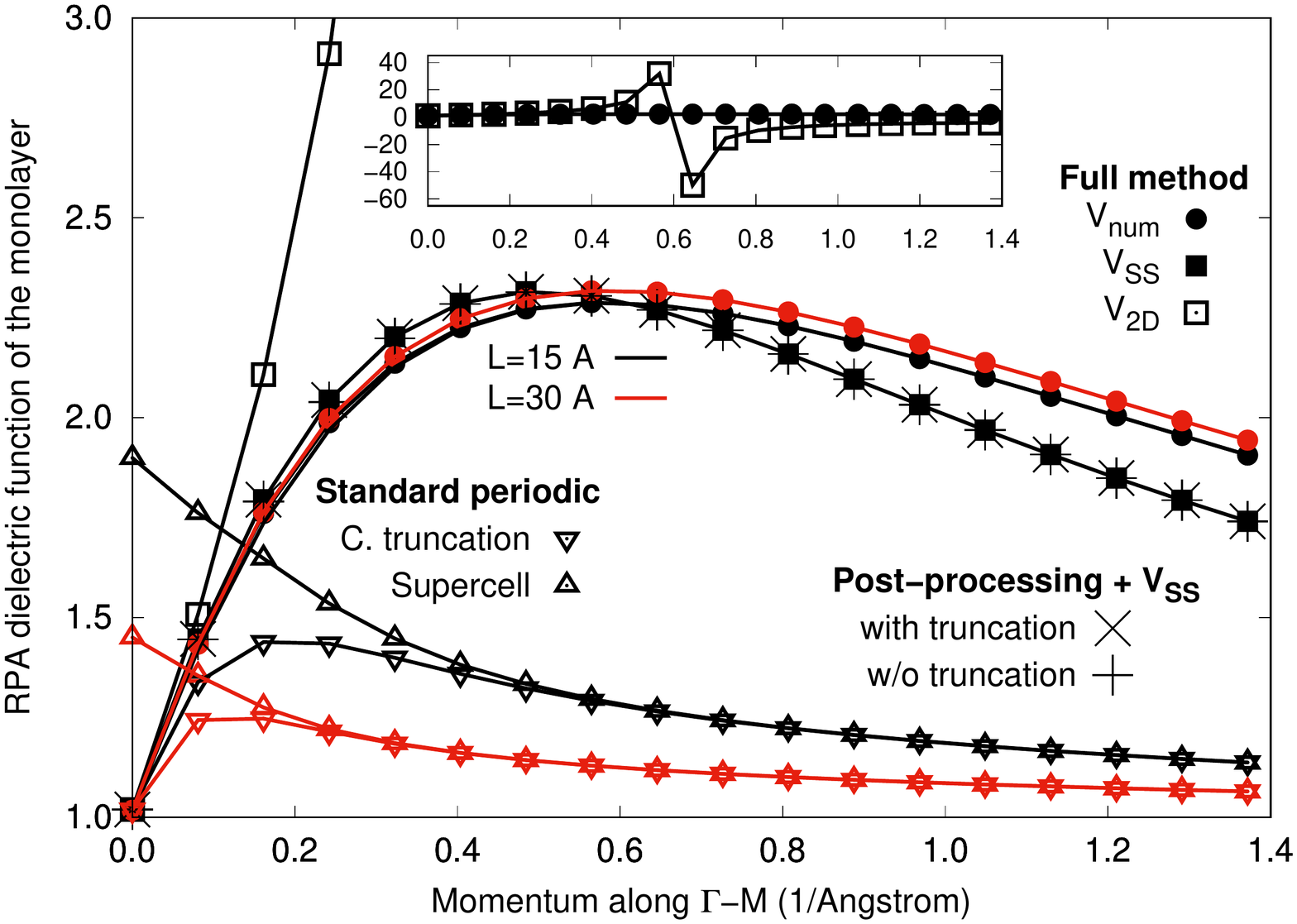}
\caption{Dielectric function of hBN single layer at finite $\qq$.
Cell height $L$=15~\AA { }(black) and $L$=30~\AA { }(red).
Standard calculations (CT = downward and SC = upward triangles), compared with~\eqref{eq:epsilon_M} evaluated with different approximations for $\mathcal{V}^{\IN}$ (circles and squares), and using~\eqref{eq:pp_cutoff} (stars) and~\eqref{eq:pp_supercell} (crosses). 
Inset: Divergence of~\eqref{eq:epsilon_M} in the 2DA.}
\label{fig:fig2}
\end{figure}

Results of the top-down strategy are plotted in Figure~\ref{fig:fig2}, which reports $\epsilon_M(\qq,\w=0)$ for $\qq=\qp$ on the $\Gamma-M$ direction of the Brillouin zone. 

%We recall that, even if the frequency variable has been dropped from all equations of this paper, all formulas are implicitly functions of $\w$.

Let us first discuss the standard methods.
In the lower part of the figure we report $\tilde{\epsilon}_M$, i.e. the output of standard ab initio calculations, in the CT scheme (downward-pointing triangles) and in the SC scheme (upward-pointing ones).
Calculations have been carried out in cells with different heights: $L=15$~\AA { }(black curves) and $L=30$~\AA { }(red ones).
In the CT calculations, the truncation appears both in $X=X^0+X^0vX$ and in $\tilde{\epsilon}_M = 1+vX$ as it is the most common implementation.
The figure clearly shows that the CT reproduces the correct long-range limit of $\epsilon_M$, while in the short range it coincides with the SC simulation.
As a consequence, the CT results have the expected shape as a function of $\qq$. 
All these are very well-known properties of the CT scheme. 

However, the figure highlights also that, except for the $\qq=0$ point, the value of $\tilde{\epsilon}_M$ does actually depend on $L$, which is still an arbitrary quantity even in this simulation scheme.
This arbitrariness can spoil, for instance, the results of quasiparticle GW corrections or excitonic calculations because it would affect the screening between the electron and the hole (on this subject see Refs~\cite{Tian2020,Qiu2016}).
Even accepting the argument of an error cancellation between the GW self-energy and the excitonic kernel (which is not perfect, anyway) one should conclude that only the position of the excitonic peak would be right, whereas both the quasiparticle gap and the excitonic binding energy would be unreliable. Instead, our method (full circles) gives results that are independent of the vacuum, the small discrepancies in the two cells being due to differences at the level of the $X^0$, and it does reproduce the expected limits of $\epsilon_M$ and has the right functional shape.
The comparison with standard calculations shows that, even when a reasonable amount of vacuum ($L=15$ \AA) is used, the CT gives results that systematically underestimate $\epsilon_M$ at finite $\qq$.

Now, let us discuss the impact of different approximations in~\eqref{eq:epsilon_M}.
Full circles correspond to results obtained with the full method, which means that we solved equation~\eqref{eq:solution_dyson} and evaluated numerically $\mathcal{V}^{\IN}$ both in~\eqref{eq:solution_voff_1} and~\eqref{eq:epsilon_M}. The empty squares correspond to the 2DA version of~\eqref{eq:epsilon_M}, which is the approximation adopted also by Nazarov~\cite{nazarov_prb2014, nazarov_njp2015}. The same $\chi_{2D}$ of the full method has been used. The inset shows clearly that neglecting completely the thickness of the layer in~\eqref{eq:epsilon_M} leads to unphysical results with a dramatic divergence at finite $\qq$.
One may account for the thickness of the slab by associating a thickness $d$ to the layer in the SSA. The results corresponding to this approximation are reported in full squares, where we employed a thickness $d=3.33$~\AA , which corresponds to the inter-layer distance in bulk hBN. At small $\qq$ the full method and the SSA one give very similar results, but they start differing at larger $\qq$ where the SSA $\epsilon_M$ decreases faster, so it systematically underestimates the dielectric function at large $\qp$. Besides this, the thickness used in the SSA is somewhat arbitrary whereas our full method is completely ab initio.

Let us now discuss the performances of the post-processing formulae~\eqref{eq:pp_cutoff} and~\eqref{eq:pp_supercell}.
Results are reported as stars and crosses respectively and they refer to calculations where the SSA has been adopted in~\eqref{eq:epsilon_M} with $d$=3.33\AA.
We preferred to exemplify the performances of the post-processing formulae with SSA instead of NUM calculations because in the latter case one should pass through the extraction of $\theta^0$, which some times may require an involved input-output handling.
It is clear that either approach leads to results that are identical to those obtained solving the full Dyson equation~\eqref{eq:solution_dyson} within the same approximation, which validates the post-processing formula in both the SL and CT framework.

Finally, it is worth stressing that equation~\eqref{eq:pp_supercell} assumes the 2DA in~\eqref{eq:solution_voff_1}.
This indicates that while this approximation is justified at the $\chi^{2D}$ level, it leads instead to dramatically wrong results when employed in~\eqref{eq:epsilon_M}, as we demonstrated above. 
This observation is related to the fundamental difference between the $\mathcal{V}^{\IN}$ appearing in~\eqref{eq:epsilon_M} and in~\eqref{eq:solution_dyson}-\eqref{eq:solution_voff_1} discussed in the previous section.

To summarize, we have derived a series of equations permitting to split the polarizability of a layered material into intra- and inter-layer contributions. An important step of this derivation is the definition of the layer-localized profile functions $\theta^n$. 
This introduces  two characteristic lengths in the derivation, the effective thickness $d$ (which is related to the localization of the basis and can be computed from first principles) and the vertical periodicity of the crystal (i.e. the height of the simulation cell) $L$. In the context of the top-down strategy, we have shown how to obtain the polarizability of an isolated sheet of matter starting from the polarizability of a periodic array of identical layers as computed with standard ab initio codes. The case of hBN has been discussed in some detail. The same formalism can be applied to the bottom-up strategy, which consists in calculating the global response of multilayers or even layered bulk materials from the responses of their constituent films.

%% The authors thank C. Giorgetti for the fruitful discussions and comments.
Funding for this work came from the European Union’s Horizon 2020 research and innovation program under grand agreement N° 881603 (Graphene Flagship core 3).

%
% 

%%%%%%%%%%%%%%%%%%%%%%%%%%%%%%%%%%%%%%%%%%%%%%%%%%%%%%%%%%%

\section{Appendix A: Derivation of the single-layer equations}

We work in the mixed space representation $\rb{\kp,z}$ where $\kp = \qp+\Gp$ is the planar component of the crystal momentum, expressed as the sum of a continuous part $\qp$ defined inside the first Brillouin zone and a reciprocal lattice vector $\Gp$.
Note that, throughout all derivations, the short wavelength contribution never changes, i.e. $\kp'=\qp+\Gp'$ and analogously $k_z' = q_z+G_z'$.  Instead, the variable $z$ spans the perpendicular direction in the real space. 
Working in the mixed space representation allows us to let physical quantities such as the atomic wavefunctions or the polarizability extend perpendicularly to the layer without assuming any periodicity in	that direction. The passage from the real-space representation $\rb{\bm{\rho},z}$ to the mix-space one is the in-plane Fourier transformation:
\begin{equation}
\lrb{\bm{\rho},z}{\kp,z'} = \frac{1}{\sqrt{\mathcal{A}}} \,e^{i\kp \cdot \bm{\rho}} \, \delta(z-z')	\nn \,,
\end{equation}
where $\mathcal{A}$ is the surface of the $xy$ plane, and $\bm{\rho}=(x,y)$ groups the in-plane real-space coordinates.
In the system of units we used, $1/4\pi\epsilon_0=\hbar=e=1$, so the real-space representation of the Coulomb interaction is $v(\rr,\rr') = 1 / | \rr - \rr' | $. In the mixed space representation, it reads:
\begin{equation}
v(\kp, z,z')  =  \frac{2\pi e^{-| \kp |\,|z-z'|}}{|\kp|} \, .
\label{eq:mixed-rep_coulomb}
\end{equation}
We now make the assumption that the electronic wavefunctions  $\phi_{i \Kp}(\rr)$ are separable:
\begin{equation}
\phi_{i \Kp}(\rr) = \varphi_{i \Kp}(\bm{\rho})\xi(z)\quad \,. 
\label{eq:factorization_wavefunctions}
\end{equation}
Note that the crystal momentum is only in-plane because we are in the framework of the isolated layer. The irreducible polarizability $\chi^0$ is calculated as usual as a sum over all independent-particle transitions. 

Owing to the factorization hypothesis~\eqref{eq:factorization_wavefunctions},the numerator of $\chi^0$ is the product of two matrix elements of the form  $\int \phi^*_{i \Kp}(\rr)e^{-i\left(\qp+\Gp\right) \cdot \bm{\rho}} \phi_{j \Kp+\qp}(\rr) \, d\bm{\rho} = \xi^2(z) M_{ij\Kp}(\qp,\Gp)$. So two profile functions $\theta^0(z)=\xi^2(z)$ can be factored out from the sum over all transitions, which leads to the expression :
\begin{equation}
 \chi^0_{\Gp\Gp'}(\qp,\w,z,z')  = \theta^0(z) \chi^0_{2D, \Gp\Gp'}(\qp,\w) \theta^0(z') \,. 
\label{eq:chi02d_mixed-rep}
\end{equation}

In evaluating numerically the profile functions $\theta^0(z)$, we observed a mild dependence on the in-plane momentum $\kp$, so we  introduce a momentum-dependent profile function $\theta^0_{\kp}(z)$ which will be used from now on. The profile functions $\theta^0_{\kp}(z)$ are defined as positive, real and even functions of $z$.
They are localized around $z=0$, which means that they are negligible for $|z|> d_{\kp}/2$ where $d_{\kp}$ is a sufficiently large distance from the slab.
This characteristic distance will be actually defined more rigorously in the layer-projected representation (see Appendix B).
Also, we take the profile functions normalized $\int \theta^0_{\kp}(z) dz = 1$, which implies that the dimension of $\theta^0_{\kp}(z)$ is the inverse of a length.

The RPA polarizability of the slab $\chi$ satisfies a Dyson equation of the form $\chi = \chi^0 + \chi^0 v \chi$.
To better appreciate its structure, let us write down the zeroth and the first order of it, while making use of definition~\eqref{eq:chi02d_mixed-rep}.
To simplify the notation, we will drop all the frequency and momentum variables, but it is understood that all polarizabilities are matrices in the $\Gp$ space and hence all the in-plane local fields are taken into account. Expanding the Dyson equation order by order we get:
\begin{equation}
\begin{split}
\lefteqn{ \chi(z,z') =\theta^0(z) \chi_{2D}^0\, \theta^0(z')  + }\\ 
& \qquad + \theta^0(z)  \chi_{2D}^0  \mathcal{V}^{\IN}  \chi_{2D}^0\,   \theta^0(z') + O(2)\,.
\end{split}
\label{eq:first-orders-chislab}
\end{equation}
In the expression above we have introduced a layer-projected Coulomb interaction
\begin{equation}
\mathcal{V}^{\IN}(\kp) = \int \int \theta^0_{\kp}(z) v(\kp,z,z') \, \theta^0_{\kp}(z') \, dz \,dz' \, ,
\label{eq:v_in}
\end{equation}
where $v(\kp,z,z')$ is defined as in~\eqref{eq:mixed-rep_coulomb}.
It is easy to convince oneself that every order in $v$ is actually localized around $z=0$ because each order is sandwiched by two profile functions left out of the integrals.
On the contrary, all the other profile functions are integrated with the Coulomb interaction $v(\kp,z_1,z_2)$ making $\mathcal{V}^{\IN}$ appear at every order.
Following this argument, one can define the 2D reducible polarizability $\chi_{2D}$ such that the RPA polarizability of the slab reads:
\begin{equation}
\chi_{\Gp\Gpp}(\qp,z,z',\w) = \theta^0_{\kp}(z)  \chi_{2D,\Gp\Gp'}(\qp,\w)  \theta^0_{\kp}(z') .
\label{eq:chi_in}
\end{equation}
The planar polarizability $\chi_{2D}$ solves the RPA 	Dyson equation:
\begin{equation}
\begin{split}
\lefteqn{ \chi_{2D\Gp\Gp'}(\qp,\w) = \chi^0_{2D\Gp\Gp'}(\qp,\w) + } \\ 
&+\sum_{\Gp''}\chi^0_{2D\Gp\Gp''}(\qp,\w)\mathcal{V}^{\IN}(\kp'')\chi_{2D\Gp''\Gp'}(\qp,\w)  \, ,
\end{split}
\label{eq:chi_2D}
\end{equation}
which depends only on the in-plane components of the crystal momentum.

\section{Appendix B: Different representations}
In the paper we work mostly with three different representations: the mixed space representation (MS) $\rb{\kp,z}$, the layer-projected representation (LP) $\rb{\kp,n}$ and the standard reciprocal space representation (RS) $\rb{\kp,k_z}$.
Many results require to move from one representation to the other.
In defining the LP basis, we project it on the MS basis introduced before. 
We can make the choice of taking the LP basis functions proportional to the profile functions $\theta^0_{\kp}$ introduced above.  
If we do this, and we note that $\theta^0_{\kp}$ has the dimension of an inverse length because of its normalization in space, then we can define  
\begin{equation}
\lrb{\kp,z}{\kp',n} =  \sqrt{d_{\kp}}\theta^n_{\kp}(z) \delta_{\Gp \Gp'} \, , \label{eq:localized-basis}
\end{equation}
where $n \in \mathbb{Z}$ is the index of the layer and the basis functions $\theta^n_{\kp}(z):=\theta^0_{\kp}(z-nL)$ satisfy the orthonormality relations:
\begin{eqnarray}
1 &=& \int \theta^n_{\kp}(z) dz \quad \text{ and } \label{eq:normal-basis} \\
 \delta_{np} &=&   d_{\kp} \int\theta^n_{\kp}(z)\theta^p_{\kp}(z)dz  \label{eq:orthogonal-basis} \,.
\end{eqnarray}
With these definitions we create an orthonormal basis of layer-localized functions, in analogy to the tight-binding formalism where site-localized atomic wavefunctions constitute the basis for crystal properties. 
Besides, the last equation is also the definition of a characteristic length $d_{\kp}$, which can be associate to an effective polarizable thickness of the layer.

Many ab initio codes assume a periodicity of the simulation cell in all directions, so they employ the RS representation $\rb{\kp,k_z}$.
Let us now evaluate its projection on the other two representations introduced above.
Its projection on the MS is a simple Fourier transformation on the $z$ variable:
\begin{equation}
\lrb{\kp,z}{\kp',k_z} = \frac{1}{\sqrt{NL}}e^{i k_z  z}\delta_{\Gp \Gp'}
\label{eq:reciprocal_to_real}
\end{equation}
where $N$ is the number of layers, so $NL$ is the vertical size of the crystal.
Its projection on the LP is:
\begin{align}
\lrb{\kp,n}{\kp',k_z} &= \sum_{\Gp''} \int \lrb{\kp,n}{\kp'',z} \lrb{\kp'',z}{\kp',k_z} dz \nn \\
& = \sqrt{\frac{d_{\kp}}{NL}} \delta_{\Gp\Gp'} \int \theta^n_{\kp}(z) e^{i k_z z} dz \nn \\
& = \sqrt{\frac{d_{\kp}}{NL}} \delta_{\Gp\Gp'} e^{i \qz nL} \vartheta_{\kp}(k_z) \,. \label{eq:projection_layer-recip3d}
\end{align}
In the last passage we used the definition of $\theta^n_{\kp}(z)$ and used the fact that $e^{i \Gz nL}   = 1$ for all $n \in \mathbb{Z}$. This defines $\vartheta$ as the  Fourier transform of the $n=0$ profile function:
\begin{equation}
\theta^0_{\kp}(z) =\frac{1}{2\pi} \int \vartheta_{\kp}(k_z) e^{i k_z z} dk_z \approx \frac{1}{L}\sum_{G_z} \vartheta_{\kp}(G_z) e^{i G_z z}\, ,
\end{equation}
which is how the Fourier transform is actually calculated using standard implementations of the discrete Fourier transform.
This discrete approximation is justified under the assumption that $L$ is large.
To conclude this part, it is useful to evaluate the Coulomb interaction in the LP representation:
\begin{align}
\lefteqn{ v_{nm}(\kp) = \lb{\kp,n} v \rb{\kp,m}  =} \nn \\ 
& = \sum_{\Gp'} \int \int \lrb{\kp,n}{\kp',z}v(\kp',z,z') \lrbb{\kp',z}{\kp,m} dz \, dz' \nn \\ 
& = d_{\kp} \int \int \theta^n_{\kp}(z)  v(\kp,z,z') \theta^m_{\kp}(z') dz \, dz'. \label{eq:vnm}
\end{align}
In the case $m=n$ it easy to show, through some simple changes of variables, that;
\begin{equation}
v_{nn}(\kp) = d_{\kp} \int \int \theta^0(z) v(\kp,z,z')\theta^0(z') dz \, dz'\,.
\end{equation}
Comparing this result with definition~\eqref{eq:v_in} and remembering that $\hat{v}_{nm}(\kp) := v_{nm}(\kp)/d_{\kp}$, one demonstrates that $\hat{v}_{nn}(\kp) \equiv \mathcal{V}^{\IN}(\kp)$.

\section{Appendix C: Rewriting in  reciprocal space}

Our main result is the Dyson equation describing the array polarizability $\hat{\chi}$ as generated by the inter-layer interactions between the single layers, namely :
\begin{widetext}
\begin{equation}
\hat{\chi}_{\Gp\Gp'nm}(\qp,\w) = \chi_{2D,\Gp\Gp'nm}(\qp,\w) + \sum_{\Gp'', p, s} \chi_{2D,\Gp\Gp''np}(\qp,\w) \mathcal{V}^{\OFF}_{ps}(\qp+\Gp'')\hat{\chi}_{\Gp''\Gp'sm}(\qp,\w) \,.
\label{eq:dyson_layer_long}
\end{equation}
\end{widetext}
In the latter equation, $\chi_{2D,\Gp\Gp'nm}$ is defined as the solution of the Dyson equation: 
\begin{equation}
\chi_{2D,nm} = \hat{\chi}^0_{nm}+\sum_{pq}\hat{\chi}^0_{np}\mathcal{V}^{\IN}_{pq} \, \chi_{2D,qm}
\label{eq:2d-dyson}
\end{equation}
resulting from the splitting of the Coulomb interaction into an intra-layer contribution  $\mathcal{V}^{\IN}_{nm}=\delta_{nm}\mathcal{V}^{\IN}$ and an inter-layer one.
We also recall that $\hat{\chi}^0$ is a sum over periodically repeated single-layer contributions, so $\hat{\chi}^0_{nm}=\delta_{nm}\chi^0_{2D}$.
This observation, together with definition of $\mathcal{V}^{\IN}_{nm}$, allow us to simplify equation~\eqref{eq:2d-dyson} as:
\begin{equation}
\chi_{2D,nm}  = \delta_{nm}\left[ \chi^0_{2D} + \chi^0_{2D} \mathcal{V}^{\IN} \chi_{2D} \right]\, ,
\end{equation}
which then becomes diagonal in the layer-projected basis, and actually layer-independent.

Let us now turn back to equation~\eqref{eq:dyson_layer_long}.
Its solution is solved by inverting the Dyson equation, which gives:
\begin{equation}
\begin{split}
\lefteqn{ \hat{\chi}_{\Gp\Gp',mn}(\qp,\w)   =} \\ 
& \qquad = \sum_{\Gp''} \mathcal{M}^{-1}_{\Gp\Gp''nm}(\qp,\w)\chi_{2D\Gp''\Gp'}(\qp,\w)   \, ,
\end{split}
\label{eq:solution1}
\end{equation}
with
\begin{equation}
\begin{split}
\lefteqn{\mathcal{M}_{\Gp\Gp'nm}(\qp,\w) = } \\
& = \delta_{\Gp\Gp'}\delta_{nm} - \chi_{2D\Gp\Gp'}(\qp,\w)\mathcal{V}^{\OFF}_{nm}(\qp+\Gp')   \, .
\end{split}
\label{eq:solution2}
\end{equation}
To make a connection between this result and the output of a standard ab initio simulation $X_{\GG\GG'}(\qq,\w)$, we shall switch to the reciprocal-space representation $\rb{\kp,k_z}$: 
\begin{align*}
\lefteqn{ X_{\GG,\GG'}(\qq) = \lrbbb{\kp, k_z}{X}{\kp', k_z'} = } \\
& = \sum_{\Gp''\Gp'''}\sum_{m,n} \lrb{\kp, k_z}{\kp'', m}X_{\kp''\kp''',mn}\lrb{\kp''', n}{\kp', k_z'}\,.
\end{align*}
Using the projection~\eqref{eq:projection_layer-recip3d} and remembering that :
\begin{equation*}
X_{\kp\kp'mn}=\sqrt{d_{\kp}d_{\kp'}}\hat{\chi}_{\kp\kp'mn}\,,
\end{equation*}
one finally arrives to:
\begin{equation}
X_{\GG,\GG'}(\qq) = \frac{1}{L}\vartheta_{\kp}(k_z) \hat{\chi}_{\Gp\Gp'}(\qp,\qz) \vartheta_{\kp'}(k_z')\,,
\label{eq:chi_in_reciprocal}
\end{equation}
where we have introduced the definition: 
\begin{equation}
\begin{split}
\lefteqn{ \hat{\chi}_{\Gp\Gp'}(\qp,\qz) = }\\
&\qquad=\frac{1}{N} \sum_{m,n} e^{-iq_zmL} \hat{\chi}_{\Gp\Gp',mn}(\qp)e^{iq_znL}\,.
\end{split}
\end{equation}
The latter quantity is basically a discrete Fourier transform of~\eqref{eq:solution1}. 
Since $\chi_{2D}$ is layer-independent, the solutions~\eqref{eq:solution1} and~\eqref{eq:solution2}  projects simply as:
\begin{equation}
\begin{split}
\lefteqn{ \hat{\chi}_{\Gp\Gp'}(\qp,\qz,\w)   = } \\ 
&\qquad = \sum_{\Gp''} \mathcal{M}^{-1}_{\Gp\Gp''}(\qp,\qz,\w)\chi_{2D\Gp''\Gp'}(\qp,\w) \label{eq:solution3}
\end{split}
\end{equation}
and
\begin{equation}
\begin{split}
\lefteqn{ \mathcal{M}_{\Gp\Gp'}(\qp,\qz,\w) = } \\ 
& \quad = \delta_{\Gp\Gp'} - \chi_{2D\Gp\Gp'}(\qp,\w)\mathcal{V}^{\OFF}(\qp+\Gp,q_z) \label{eq:solution4}\,.
\end{split}
\end{equation}
The last step to complete the connection with the reciprocal-space representation is the expression of $\mathcal{V}^{\OFF}(\qp+\Gp,q_z)$.
Remembering its definition in the layer-projected basis (it vanishes for $m=n$), one writes
\begin{equation}
\mathcal{V}^{\OFF}(\kp,\qz) = \frac{1}{N}\sum_{m} \sum_{n\neq m } e^{-iq_z mL} \hat{v}_{mn}(\kp)e^{iq_z nL} \nn  \;,
\end{equation}
and $\hat{v}_{nm}(\kp)=d_{\kp}^{-1}v_{nm}(\kp)$ where $v_{nm}(\kp)$ is defined in~\eqref{eq:vnm}.
Because of the localization of the $\theta^n$ functions, if $n<m$ then $|z-z'|>0$ and vice versa. 
It is therefore convenient to split $\sum_{n\neq m} = \sum_{n<m} + \sum_{n>m}$. 
Next, inside the first sum, we change variables $z-mL=\zeta$ and $z'-nL=\zeta'$, while inside the second one, we put $z-mL=-\zeta$ and $z'-nL=-\zeta'$.
Rearranging all terms, using the parity of the $\theta^0(z)$ functions and the identity $\hat{v}_{00}=\mathcal{V}^{\IN}$, we get
\begin{equation}
\begin{split}
&\mathcal{V}^{\OFF}(\kp,\qz) =\\
& \qquad = \mathcal{V}^{\IN}(\kp)  \frac{1}{N}\sum_{m} \left[ \sum_{n<m } e^{-(iq_z + |\kp|)(m-n)L} + \right.\\
&\left. \qquad \quad +  \sum_{n>m} e^{-(iq_z-|\kp|)(m-n)L} \right]\,.
\end{split}
\nn
\end{equation}
Since both sums depend only on the difference $p=m-n$, we can change the index and cast them in the form $\sum_{p=1}^{\infty} e^{-(|\kp| \pm iq_z)Lp} = \left(  e^{(|\kp| \pm i q_z)L} - 1 \right)^{-1} $. The resulting expression for the inter-layer Coulomb interaction reads
\begin{equation}
\begin{split}
&\mathcal{V}^{\OFF}(\qp+\Gp,\qz) = \mathcal{V}^{\IN}(\qp+\Gp) \times \\
&\times \left( \frac{1}{ e^{(|\qp+\Gp| + i q_z)L} - 1 } + \frac{1}{ e^{(|\qp+\Gp| - i q_z)L} - 1 } \right)\,,
\end{split}
\label{eq:solution_voff}
\end{equation}
which, evaluated at $q_z=0$, gives the result reported in the main text.

\section{Appendix D: The dielectric function}
\subsection{In the layer-projected representation}
The definition of the microscopic dielectric function in terms of the total and external fields is:
\begin{equation}
\begin{split}
\lefteqn{U^\text{tot}_{\Gp}(\qp,\w,z) =} \\
& = \sum_{\Gp'}\int \epsilon^{-1}_{\Gp\Gp'}(\qp,\w,z,z')U^\text{ext}_{\Gp'}(\qp,\w,z') dz'  \, .
\end{split}
\label{eq:umicro}
\end{equation}
We split now the total potential  into  an external and an induced part, and assume the latter to be the classical Hartree potential, then the equation above leads to the definition:
\begin{equation}
\begin{split}
\lefteqn{ \epsilon^{-1}_{\Gp\Gp'}(\qp,\w,z,z') = \delta_{\Gp\Gp'}\delta(z-z') + } \\ 
& \qquad  + \int v(\kp,z,z'')\chi_{\Gp\Gp'}(\qp,\w,z'',z')dz''\; ,
\end{split}
\label{eq:dielfunct}
\end{equation}
where $\chi$ is defined as in~\eqref{eq:chi_in}.

We introduce now the macroscopic average of the fields, i.e. the $\Gp=0$ component projected on the single-layer. 
To this aim, we refer to the projection~\eqref{eq:localized-basis}, where we take $n=0$ and $N=1$ since we are working on the single-layer framework. 
Then, for $\alpha=$(tot) or (ext),
\begin{equation}
U^{\alpha}_{M}(\qp,\w) = \sqrt{d_{\qp}} \int \theta^0_{\qp}(z)U^{\alpha}_0(\qp,\w,z)dz  \, ,
\label{eq:umacro}
\end{equation}
and we define $U^{\text{tot}}_{M}(\qp,\w) = U^{\text{ext}}_{M}(\qp,\w)/\epsilon_M(\qp,\w)$. We can now plug~\eqref{eq:umicro} into~\eqref{eq:umacro} and assume that $U^\text{ext}$ is macroscopic, which means that the only non-vanishing term is the $\Gp=0$ one.
From the definitions~\eqref{eq:dielfunct} and~\eqref{eq:chi_in}, we get the relation:
\begin{widetext}
\begin{equation}
 U^\text{tot}_{M}(\qp,\w) = \sqrt{d_{\qp}}\int \int \theta^0_{\qp}(z) \left[ \delta(z-z')  + \int v(\qp,z,z')\theta^0_{\qp}(z'')\chi_{2D,00}(\qp,\w)\theta^0_{\qp}(z')dz''\right] U^{\text{ext}}_0(\qp,\w) dz' dz  \, ,
\end{equation}
\end{widetext}
which eventually leads to the single layer macroscopic dielectric function:
\begin{equation}
\frac{1}{\epsilon_M(\qp,\w)} = 1 + \mathcal{V}^{\IN}(\qp)\chi_{2D,00}(\qp,\w) \, ;
\label{eq:eps_macro}
\end{equation}

\subsection{In the reciprocal-space representation}
Let us now repeat the derivation in the $\rb{\kp,k_z}$ representation.
This will allow us to show that our approach reproduces some results obtained by other authors in the sharp slab approximation \cite{latini_prb2015}, but actually permits to go beyond them generalizing their method. Similarly to what done xbefore, we use~\eqref{eq:projection_layer-recip3d} with $n=0$ and $N=1$.
Furthermore we set from the beginning $\Gp=0$, we assume $U^\text{ext}$ to be macroscopic, and we drop the frequency variable for notational convenience.
We get:
\begin{eqnarray}
U^\text{tot}_M(\qp) &=& \sqrt{\frac{d_{\qp}}{L}} \int \sum_{\Gz} \vartheta_{\qp}(k_z) U^\text{tot}_{(0,G_z)}(\qp,q_z) d\qz  \nn \\
U^\text{ext}_M(\qp) &=&  \sqrt{\frac{d_{\qp}}{L}} \int \vartheta_{\qp}(q_z) U^\text{ext}_{(0,0)}(\qp,\qz)  d\qz   \nn \,  .
\end{eqnarray}
We can do the additional assumption that the integrands depend weakly on $\qz$ and hence fix $\qz=0$ and replace $\int f(q_z) d\qz \approx L f(0) /2\pi$ in both expressions.
We get the two definitions:
\begin{eqnarray}
U^\text{tot}_M(\qp) & \approx & \frac{\sqrt{d_{\qp} L}}{2\pi} \sum_{\Gz} \vartheta_{\qp}(G_z) U^\text{tot}_{(0,G_z)}(\qp,0)  \label{eq:utot_macro_bis} \\
U^\text{ext}_M(\qp) & \approx &  \frac{\sqrt{d_{\qp} L}}{2\pi} U^\text{ext}_{(0,0)}(\qp,0)    \label{eq:uext_macro_bis} \, ,
\end{eqnarray}
where we have used the fact that $\vartheta_{\qp}(0)=1 \, \forall \qp$ as a consequence of its normalization in real space.

In the $\rb{\kp,k_z}$  representation, the  microscopic dielectric function is defined according to the relation $ 
U^\text{tot}_{\GG}(\qq)  = \sum_{\GG'} \epsilon^{-1}_{\GG\GG'}(\qq) U^{\text{ext}}_{\GG'}(\qq) $.
Inserting this definition into~\eqref{eq:utot_macro_bis}, using~\eqref{eq:uext_macro_bis}, and using the fact that $U^{\text{ext}}$ is macroscopic, one gets:
\begin{equation}
U^\text{tot}_M(\qp)  = \sum_{\Gz} \vartheta_{\qp}(G_z) \epsilon^{-1}_{(0,\Gz),(0,0)}(\qp) U^{\text{ext}}_M(\qp) \nn  \, ,
\end{equation}
from which one defines:
\begin{equation}
\frac{1}{\epsilon_M(\qp)} = \sum_{\Gz} \vartheta_{\qp}(G_z) \epsilon^{-1}_{(0,\Gz),(0,0)}(\qp) \,.
\label{eq:latini_generalized}
\end{equation}
This result is a generalization of the \emph{quasi 2D} dielectric function derived by Latini et al. \cite{latini_prb2015}.
Their result can be retrieved within the \emph{sharp slab} approximation, which in the reciprocal space gives $\vartheta_{\qp}(G_z) = (\Gz \frac{d}{2} )^{-1} \sin(\Gz\frac{d}{2})$.

Let us now show that~\eqref{eq:latini_generalized} leads to the same expression we derived in the main text.
First of all, let us recall that the dielectric function in the $\rb{\kp,k_z}$ representation reads:
\begin{equation}
\epsilon^{-1}_{\GG\GG'}(\qq,\w) = \delta_{\GG\GG'} + v(\qq+\GG)\chi_{\GG\GG'}(\qq,\w) \,
\label{eq:epsilon}
\end{equation}
and that the single layer polarizability is:
\begin{equation}
\chi_{\GG\GG'}(\qq,\w) = \frac{1}{L}\vartheta_{\kp}(k_z)\chi_{2D, \Gp \Gp'}(\qp,\w)\vartheta_{\kp}(k_z')  \, ,
\label{eq:chi}
\end{equation}
in analogy to~\eqref{eq:chi_in_reciprocal}.
Actually, in equation~\eqref{eq:chi_in_reciprocal}, $L$ is the period of the periodic array, and $NL$ the height of the crystal.
In expression~\eqref{eq:chi} $L$ has to be interpreted as the size of the box of integration, and $N=1$.
However, in practice, it still corresponds to the height of the simulation cell. 
In any case, at the end of the calculation this factor will cancel out, in agreement with the physical requirement that the single-layer dielectric function does not depend on the size of the simulation cell.

Using~\eqref{eq:epsilon} and~\eqref{eq:chi}, and remembering that $\vartheta_{\kp}(0)=1$, equation~\eqref{eq:latini_generalized} becomes:
\begin{equation}
\frac{1}{\epsilon_M(\qp,\w)}  =  1 +  \frac{\chi_{2D\,00}(\qp,\w)}{L}  \sum_{\Gz} \vartheta_{\qp}^2(\Gz) v(\qp,G_z) \,.\nn
\end{equation}

\begin{widetext}
Finally, to complete the demonstration, we shall evaluate the term $\sum_{\Gz}$.
This is done by projecting successively on the different representations (remember $N=1$).
We get:

\begin{eqnarray}
\lefteqn{\sum_{\Gz} \vartheta_{\qp}^2(\Gz) v(\qp,G_z)  = }\nn\\
& =& \frac{L}{d_{\qp}}  \sum_{\Gz\Gz'} \lrb{\qp,n=0}{\qp,G_z} \lb{\qp,G_z}v \rb{\qp,G_z'} \lrbb{\qp,G_z'}{\qp,n=0} \nn \\
& =&  \frac{L}{d_{\qp}}  \sum_{\Gz\Gz'} \int \int \lrb{\qp,0}{\qp,z} \lrb{\qp,z}{\qp,G_z} \lb{\qp,G_z}v \rb{\qp,G_z'} \lrb{\qp,G_z'}{\qp,z'}\lrb{\qp,z'}{\qp,0} dz dz'  \, .\nn 
\end{eqnarray}
We use definitions~\eqref{eq:localized-basis}, \eqref{eq:reciprocal_to_real} and \eqref{eq:projection_layer-recip3d} and arrive to the expression;
\begin{equation}
\sum_{\Gz} \vartheta_{\qp}^2(\Gz) v(\qp,G_z)  = \sum_{\Gz\Gz'} \int \int \theta^0_{\qp}(z)   e^{i\Gz z}\frac{4\pi }{| \qp + \hat{z}\Gz |^2} \delta_{\Gz \Gz'}  e^{- i\Gz'z'}\theta^0_{\qp}(z') dz\, dz' \,.
\end{equation}
If we notice that $v(\qp,z,z') = \frac{1}{L} \sum_{\Gz} v(\qp+\hat{z}\Gz)e^{i\Gz (z-z')}$ is a Fourier transform at $\qz=0$ and $N=1$, then we are left with L times the definition~\eqref{eq:v_in}, so the result becomes:
\begin{equation}
\sum_{\Gz} \vartheta_{\qp}^2(\Gz) v(\qp,G_z)  = L \mathcal{V}^{\IN}(\qp)\, ,
\end{equation}
which completes the demonstration.

\bigskip
\bigskip
\bigskip
\bigskip
\bigskip
\bigskip

\end{widetext}

\section{Appendix E: Computational details}

In this section we give the details of the simulations of the hBN monolayer.

All ground state calculations have been carried out with the ABINIT simulation package \cite{abinit}.
The parameters of the simulation cell are $a=2.5$~\AA { }for the in-plane lattice parameter and $L=15$~\AA { }or $L=30$~\AA { }for the height of the simulation cell.
In all cases, the local density approximation (LDA) was used to model the exchange-correlation potential of the Kohn-Sham Hamiltonian and the cutoff energy defining the basis set was 30 Ha.
The k-point grid used to sample the electronic density was $15 \times 15 \times 1$, instead for the wavefunction calculations, necessary to compute $X^0$, a denser grid of $36 \times 36$ k-points in the $(x,y)$-plane was employed.
Both grids were centered in $\Gamma$. 
No Coulomb truncation has been used in the ground state calculations.

The wavefunctions have been successively passed to the DP \cite{dp} simulation package with which we computed only the irreducible polarizability  $X^0$ of the periodic array.
The sum over all transitions included 30 bands in the $L=15$~\AA { }calculations, and 60 in the $L=30$~\AA { }ones.
The cutoff energy for the representation of the matrix elements was 400~eV, and it was 100~eV for the dimension of the $X_{\GG\GG'}$.
No scissor operator has been applied, so the resulting gap was the LDA one.

All the RPA algorithms connecting $X^0$ to $X$ and $\tilde{\epsilon}_M$, with and without Coulomb truncation, as well as all the manipulations proper to our method have been implemented in a home-made code which takes $X^0$ in input.


%%% \bibliography{biblio_v5}

\begin{thebibliography}{19}%
\makeatletter
\providecommand \@ifxundefined [1]{%
 \@ifx{#1\undefined}
}%
\providecommand \@ifnum [1]{%
 \ifnum #1\expandafter \@firstoftwo
 \else \expandafter \@secondoftwo
 \fi
}%
\providecommand \@ifx [1]{%
 \ifx #1\expandafter \@firstoftwo
 \else \expandafter \@secondoftwo
 \fi
}%
\providecommand \natexlab [1]{#1}%
\providecommand \enquote  [1]{``#1''}%
\providecommand \bibnamefont  [1]{#1}%
\providecommand \bibfnamefont [1]{#1}%
\providecommand \citenamefont [1]{#1}%
\providecommand \href@noop [0]{\@secondoftwo}%
\providecommand \href [0]{\begingroup \@sanitize@url \@href}%
\providecommand \@href[1]{\@@startlink{#1}\@@href}%
\providecommand \@@href[1]{\endgroup#1\@@endlink}%
\providecommand \@sanitize@url [0]{\catcode `\\12\catcode `\$12\catcode
  `\&12\catcode `\#12\catcode `\^12\catcode `\_12\catcode `\%12\relax}%
\providecommand \@@startlink[1]{}%
\providecommand \@@endlink[0]{}%
\providecommand \url  [0]{\begingroup\@sanitize@url \@url }%
\providecommand \@url [1]{\endgroup\@href {#1}{\urlprefix }}%
\providecommand \urlprefix  [0]{URL }%
\providecommand \Eprint [0]{\href }%
\providecommand \doibase [0]{http://dx.doi.org/}%
\providecommand \selectlanguage [0]{\@gobble}%
\providecommand \bibinfo  [0]{\@secondoftwo}%
\providecommand \bibfield  [0]{\@secondoftwo}%
\providecommand \translation [1]{[#1]}%
\providecommand \BibitemOpen [0]{}%
\providecommand \bibitemStop [0]{}%
\providecommand \bibitemNoStop [0]{.\EOS\space}%
\providecommand \EOS [0]{\spacefactor3000\relax}%
\providecommand \BibitemShut  [1]{\csname bibitem#1\endcsname}%
\let\auto@bib@innerbib\@empty
%</preamble>
\bibitem [{\citenamefont {Cudazzo}\ \emph {et~al.}(2011)\citenamefont
  {Cudazzo}, \citenamefont {Tokatly},\ and\ \citenamefont
  {Rubio}}]{Cudazzo2011}%
  \BibitemOpen
  \bibfield  {author} {\bibinfo {author} {\bibfnamefont {P.}~\bibnamefont
  {Cudazzo}}, \bibinfo {author} {\bibfnamefont {I.~V.}\ \bibnamefont
  {Tokatly}}, \ and\ \bibinfo {author} {\bibfnamefont {A.}~\bibnamefont
  {Rubio}},\ }\href {\doibase 10.1103/PhysRevB.84.085406} {\bibfield  {journal}
  {\bibinfo  {journal} {Physical Review B - Condensed Matter and Materials
  Physics}\ }\textbf {\bibinfo {volume} {84}},\ \bibinfo {pages} {085406}
  (\bibinfo {year} {2011})}\BibitemShut {NoStop}%
\bibitem [{\citenamefont {Sponza}\ \emph {et~al.}(2016)\citenamefont {Sponza},
  \citenamefont {Goniakowski},\ and\ \citenamefont {Noguera}}]{Sponza2016}%
  \BibitemOpen
  \bibfield  {author} {\bibinfo {author} {\bibfnamefont {L.}~\bibnamefont
  {Sponza}}, \bibinfo {author} {\bibfnamefont {J.}~\bibnamefont {Goniakowski}},
  \ and\ \bibinfo {author} {\bibfnamefont {C.}~\bibnamefont {Noguera}},\ }\href
  {\doibase 10.1103/PhysRevB.93.195435} {\bibfield  {journal} {\bibinfo
  {journal} {Physical Review B}\ }\textbf {\bibinfo {volume} {93}},\ \bibinfo
  {pages} {195435} (\bibinfo {year} {2016})}\BibitemShut {NoStop}%
\bibitem [{\citenamefont {Tian}\ \emph {et~al.}(2020)\citenamefont {Tian},
  \citenamefont {Scullion}, \citenamefont {Hughes}, \citenamefont {Li},
  \citenamefont {Shih}, \citenamefont {Coleman}, \citenamefont {Chhowalla},\
  and\ \citenamefont {Santos}}]{Tian2020}%
  \BibitemOpen
  \bibfield  {author} {\bibinfo {author} {\bibfnamefont {T.}~\bibnamefont
  {Tian}}, \bibinfo {author} {\bibfnamefont {D.}~\bibnamefont {Scullion}},
  \bibinfo {author} {\bibfnamefont {D.}~\bibnamefont {Hughes}}, \bibinfo
  {author} {\bibfnamefont {L.~H.}\ \bibnamefont {Li}}, \bibinfo {author}
  {\bibfnamefont {C.-J.}\ \bibnamefont {Shih}}, \bibinfo {author}
  {\bibfnamefont {J.}~\bibnamefont {Coleman}}, \bibinfo {author} {\bibfnamefont
  {M.}~\bibnamefont {Chhowalla}}, \ and\ \bibinfo {author} {\bibfnamefont
  {E.~J.~G.}\ \bibnamefont {Santos}},\ }\href {\doibase
  10.1021/acs.nanolett.9b02982} {\bibfield  {journal} {\bibinfo  {journal}
  {Nano Letters}\ }\textbf {\bibinfo {volume} {20}},\ \bibinfo {pages} {841}
  (\bibinfo {year} {2020})}\BibitemShut {NoStop}%
\bibitem [{\citenamefont {Ismail-Beigi}(2006)}]{ismail-beigi_prb2006}%
  \BibitemOpen
  \bibfield  {author} {\bibinfo {author} {\bibfnamefont {S.}~\bibnamefont
  {Ismail-Beigi}},\ }\href {\doibase 10.1103/PhysRevB.73.233103} {\bibfield
  {journal} {\bibinfo  {journal} {Physical Review B - Condensed Matter and
  Materials Physics}\ }\textbf {\bibinfo {volume} {73}},\ \bibinfo {pages} {1}
  (\bibinfo {year} {2006})}\BibitemShut {NoStop}%
\bibitem [{\citenamefont {Rozzi}\ \emph {et~al.}(2006)\citenamefont {Rozzi},
  \citenamefont {Varsano}, \citenamefont {Marini},\ and\ \citenamefont
  {Gross}}]{rozzi_prb2006}%
  \BibitemOpen
  \bibfield  {author} {\bibinfo {author} {\bibfnamefont {C.~A.}\ \bibnamefont
  {Rozzi}}, \bibinfo {author} {\bibfnamefont {D.}~\bibnamefont {Varsano}},
  \bibinfo {author} {\bibfnamefont {A.}~\bibnamefont {Marini}}, \ and\ \bibinfo
  {author} {\bibfnamefont {E.~K.~U.}\ \bibnamefont {Gross}},\ }\href {\doibase
  10.1103/PhysRevB.73.205119} {\bibfield  {journal} {\bibinfo  {journal}
  {Physical Review B}\ ,\ \bibinfo {pages} {1}} (\bibinfo {year}
  {2006})}\BibitemShut {NoStop}%
\bibitem [{\citenamefont {H{\"{u}}ser}\ \emph {et~al.}(2013)\citenamefont
  {H{\"{u}}ser}, \citenamefont {Olsen},\ and\ \citenamefont
  {Thygesen}}]{huser_prb2013}%
  \BibitemOpen
  \bibfield  {author} {\bibinfo {author} {\bibfnamefont {F.}~\bibnamefont
  {H{\"{u}}ser}}, \bibinfo {author} {\bibfnamefont {T.}~\bibnamefont {Olsen}},
  \ and\ \bibinfo {author} {\bibfnamefont {K.~S.}\ \bibnamefont {Thygesen}},\
  }\href {\doibase 10.1103/PhysRevB.87.235132} {\bibfield  {journal} {\bibinfo
  {journal} {Physical Review B}\ }\textbf {\bibinfo {volume} {87}},\ \bibinfo
  {pages} {235132} (\bibinfo {year} {2013})}\BibitemShut {NoStop}%
\bibitem [{\citenamefont {Cudazzo}\ \emph {et~al.}(2016)\citenamefont
  {Cudazzo}, \citenamefont {Sponza}, \citenamefont {Giorgetti}, \citenamefont
  {Reining}, \citenamefont {Sottile},\ and\ \citenamefont
  {Gatti}}]{cudazzo_prl2016}%
  \BibitemOpen
  \bibfield  {author} {\bibinfo {author} {\bibfnamefont {P.}~\bibnamefont
  {Cudazzo}}, \bibinfo {author} {\bibfnamefont {L.}~\bibnamefont {Sponza}},
  \bibinfo {author} {\bibfnamefont {C.}~\bibnamefont {Giorgetti}}, \bibinfo
  {author} {\bibfnamefont {L.}~\bibnamefont {Reining}}, \bibinfo {author}
  {\bibfnamefont {F.}~\bibnamefont {Sottile}}, \ and\ \bibinfo {author}
  {\bibfnamefont {M.}~\bibnamefont {Gatti}},\ }\href {\doibase
  10.1103/PhysRevLett.116.066803} {\bibfield  {journal} {\bibinfo  {journal}
  {Physical Review Letters}\ }\textbf {\bibinfo {volume} {116}},\ \bibinfo
  {pages} {066803} (\bibinfo {year} {2016})}\BibitemShut {NoStop}%
\bibitem [{\citenamefont {Nazarov}\ \emph {et~al.}(2014)\citenamefont
  {Nazarov}, \citenamefont {Alharbi}, \citenamefont {Fisher},\ and\
  \citenamefont {Kais}}]{nazarov_prb2014}%
  \BibitemOpen
  \bibfield  {author} {\bibinfo {author} {\bibfnamefont {V.~U.}\ \bibnamefont
  {Nazarov}}, \bibinfo {author} {\bibfnamefont {F.}~\bibnamefont {Alharbi}},
  \bibinfo {author} {\bibfnamefont {T.~S.}\ \bibnamefont {Fisher}}, \ and\
  \bibinfo {author} {\bibfnamefont {S.}~\bibnamefont {Kais}},\ }\href {\doibase
  10.1103/PhysRevB.89.195423} {\bibfield  {journal} {\bibinfo  {journal}
  {Physical Review B - Condensed Matter and Materials Physics}\ }\textbf
  {\bibinfo {volume} {89}},\ \bibinfo {pages} {195423} (\bibinfo {year}
  {2014})}\BibitemShut {NoStop}%
\bibitem [{\citenamefont {Nazarov}(2015)}]{nazarov_njp2015}%
  \BibitemOpen
  \bibfield  {author} {\bibinfo {author} {\bibfnamefont {V.~U.}\ \bibnamefont
  {Nazarov}},\ }\href {\doibase 10.1088/1367-2630/17/7/073018} {\bibfield
  {journal} {\bibinfo  {journal} {New Journal of Physics}\ }\textbf {\bibinfo
  {volume} {17}},\ \bibinfo {pages} {73018} (\bibinfo {year}
  {2015})}\BibitemShut {NoStop}%
\bibitem [{\citenamefont {Latini}\ \emph {et~al.}(2015)\citenamefont {Latini},
  \citenamefont {Olsen},\ and\ \citenamefont {Thygesen}}]{latini_prb2015}%
  \BibitemOpen
  \bibfield  {author} {\bibinfo {author} {\bibfnamefont {S.}~\bibnamefont
  {Latini}}, \bibinfo {author} {\bibfnamefont {T.}~\bibnamefont {Olsen}}, \
  and\ \bibinfo {author} {\bibfnamefont {K.~S.}\ \bibnamefont {Thygesen}},\
  }\href {\doibase 10.1103/PhysRevB.92.245123} {\bibfield  {journal} {\bibinfo
  {journal} {Physical Review B}\ }\textbf {\bibinfo {volume} {92}},\ \bibinfo
  {pages} {1} (\bibinfo {year} {2015})}\BibitemShut {NoStop}%
\bibitem [{\citenamefont {Thygesen}(2017)}]{thygesen_2D2017}%
  \BibitemOpen
  \bibfield  {author} {\bibinfo {author} {\bibfnamefont {K.~S.}\ \bibnamefont
  {Thygesen}},\ }\href {\doibase 10.1088/2053-1583/aa6432} {\bibfield
  {journal} {\bibinfo  {journal} {2D Materials}\ }\textbf {\bibinfo {volume}
  {4}},\ \bibinfo {pages} {022004} (\bibinfo {year} {2017})}\BibitemShut
  {NoStop}%
\bibitem [{\citenamefont {Meckbach}\ \emph {et~al.}(2018)\citenamefont
  {Meckbach}, \citenamefont {Stroucken},\ and\ \citenamefont
  {Koch}}]{Meckbach2018a}%
  \BibitemOpen
  \bibfield  {author} {\bibinfo {author} {\bibfnamefont {L.}~\bibnamefont
  {Meckbach}}, \bibinfo {author} {\bibfnamefont {T.}~\bibnamefont {Stroucken}},
  \ and\ \bibinfo {author} {\bibfnamefont {S.~W.}\ \bibnamefont {Koch}},\
  }\href {\doibase 10.1103/PhysRevB.97.035425} {\bibfield  {journal} {\bibinfo
  {journal} {Physical Review B}\ }\textbf {\bibinfo {volume} {97}},\ \bibinfo
  {pages} {035425} (\bibinfo {year} {2018})}\BibitemShut {NoStop}%
\bibitem [{\citenamefont {Gjerding}\ \emph {et~al.}(2020)\citenamefont
  {Gjerding}, \citenamefont {Cavalcante}, \citenamefont {Chaves},\ and\
  \citenamefont {Thygesen}}]{Gjerding2020}%
  \BibitemOpen
  \bibfield  {author} {\bibinfo {author} {\bibfnamefont {M.~N.}\ \bibnamefont
  {Gjerding}}, \bibinfo {author} {\bibfnamefont {L.~S.~R.}\ \bibnamefont
  {Cavalcante}}, \bibinfo {author} {\bibfnamefont {A.}~\bibnamefont {Chaves}},
  \ and\ \bibinfo {author} {\bibfnamefont {K.~S.}\ \bibnamefont {Thygesen}},\
  }\href {\doibase 10.1021/acs.jpcc.0c01635} {\bibfield  {journal} {\bibinfo
  {journal} {The Journal of Physical Chemistry C}\ }\textbf {\bibinfo {volume}
  {124}},\ \bibinfo {pages} {11609} (\bibinfo {year} {2020})}\BibitemShut
  {NoStop}%
\bibitem [{\citenamefont {Andersen}\ \emph {et~al.}(2015)\citenamefont
  {Andersen}, \citenamefont {Latini},\ and\ \citenamefont
  {Thygesen}}]{andersen_nl2015}%
  \BibitemOpen
  \bibfield  {author} {\bibinfo {author} {\bibfnamefont {K.}~\bibnamefont
  {Andersen}}, \bibinfo {author} {\bibfnamefont {S.}~\bibnamefont {Latini}}, \
  and\ \bibinfo {author} {\bibfnamefont {K.~S.}\ \bibnamefont {Thygesen}},\
  }\href {\doibase 10.1021/acs.nanolett.5b01251} {\bibfield  {journal}
  {\bibinfo  {journal} {Nano Letters}\ }\textbf {\bibinfo {volume} {15}},\
  \bibinfo {pages} {4616} (\bibinfo {year} {2015})}\BibitemShut {NoStop}%
\bibitem [{\citenamefont {Cavalcante}\ \emph {et~al.}(2018)\citenamefont
  {Cavalcante}, \citenamefont {Chaves}, \citenamefont {{Van Duppen}},
  \citenamefont {Peeters},\ and\ \citenamefont {Reichman}}]{Cavalcante2018a}%
  \BibitemOpen
  \bibfield  {author} {\bibinfo {author} {\bibfnamefont {L.~S.~R.}\
  \bibnamefont {Cavalcante}}, \bibinfo {author} {\bibfnamefont
  {A.}~\bibnamefont {Chaves}}, \bibinfo {author} {\bibfnamefont
  {B.}~\bibnamefont {{Van Duppen}}}, \bibinfo {author} {\bibfnamefont {F.~M.}\
  \bibnamefont {Peeters}}, \ and\ \bibinfo {author} {\bibfnamefont {D.~R.}\
  \bibnamefont {Reichman}},\ }\href {\doibase 10.1103/PhysRevB.97.125427}
  {\bibfield  {journal} {\bibinfo  {journal} {Physical Review B}\ }\textbf
  {\bibinfo {volume} {97}},\ \bibinfo {pages} {125427} (\bibinfo {year}
  {2018})}\BibitemShut {NoStop}%
\bibitem [{Note1()}]{Note1}%
  \BibitemOpen
  \bibinfo {note} {Note that in this notation, the continuous part never
  changes. So if $\protect \mathbf {k}_\parallel =\protect \mathbf
  {q}_\parallel +\protect \mathbf {G}_\parallel $, then $\protect \mathbf
  {k}_\parallel '=\protect \mathbf {q}_\parallel +\protect \mathbf
  {G}_\parallel '$ and analogously $k_z' = q_z+G_z'$.}\BibitemShut {Stop}%
\bibitem [{\citenamefont {Qiu}\ \emph {et~al.}(2016)\citenamefont {Qiu},
  \citenamefont {da~Jornada},\ and\ \citenamefont {Louie}}]{Qiu2016}%
  \BibitemOpen
  \bibfield  {author} {\bibinfo {author} {\bibfnamefont {D.~Y.}\ \bibnamefont
  {Qiu}}, \bibinfo {author} {\bibfnamefont {F.~H.}\ \bibnamefont {da~Jornada}},
  \ and\ \bibinfo {author} {\bibfnamefont {S.~G.}\ \bibnamefont {Louie}},\
  }\href {\doibase 10.1103/PhysRevB.93.235435} {\bibfield  {journal} {\bibinfo
  {journal} {Physical Review B}\ }\textbf {\bibinfo {volume} {93}},\ \bibinfo
  {pages} {235435} (\bibinfo {year} {2016})}\BibitemShut {NoStop}%
\bibitem [{\citenamefont {Gonze}\ \emph {et~al.}(2016)\citenamefont {Gonze},
  \citenamefont {Jollet}, \citenamefont {F.}, \citenamefont {Adams},
  \citenamefont {Amadon}, \citenamefont {Applencourt}, \citenamefont
  {Audouzec}, \citenamefont {Beuken}, \citenamefont {Bieder}, \citenamefont
  {Bokhanchuk}, \citenamefont {Bousquet}, \citenamefont {F.}, \citenamefont
  {Caliste}, \citenamefont {Côté}, \citenamefont {Dahm}, \citenamefont
  {Da~Pieve}, \citenamefont {Delaveau}, \citenamefont {Di~Gennaro},
  \citenamefont {Dorado}, \citenamefont {Espejo}, \citenamefont {Geneste},
  \citenamefont {Genovese}, \citenamefont {Gerossier}, \citenamefont
  {Giantomassi}, \citenamefont {Gillet}, \citenamefont {Hamann}, \citenamefont
  {He}, \citenamefont {Jomard}, \citenamefont {Laflamme~Janssen}, \citenamefont
  {Le~Roux}, \citenamefont {Levitt}, \citenamefont {Lherbier}, \citenamefont
  {Liu}, \citenamefont {Lukačević}, \citenamefont {Martin}, \citenamefont
  {Martins}, \citenamefont {Oliveira}, \citenamefont {Poncé}, \citenamefont
  {Pouillon}, \citenamefont {Rangel}, \citenamefont {Rignanese}, \citenamefont
  {Romero}, \citenamefont {Rousseau}, \citenamefont {Rubel}, \citenamefont
  {Shukri}, \citenamefont {M}, \citenamefont {Torrent}, \citenamefont
  {Van~Setten}, \citenamefont {Van~Troeye}, \citenamefont {Verstraete},
  \citenamefont {Waroquiers}, \citenamefont {Wiktor}, \citenamefont {Xu},
  \citenamefont {Zhou},\ and\ \citenamefont {Zwanziger}}]{abinit}%
  \BibitemOpen
  \bibfield  {author} {\bibinfo {author} {\bibfnamefont {X.}~\bibnamefont
  {Gonze}}, \bibinfo {author} {\bibfnamefont {F.}~\bibnamefont {Jollet}},
  \bibinfo {author} {\bibfnamefont {A.~A.}\ \bibnamefont {F.}}, \bibinfo
  {author} {\bibfnamefont {D.}~\bibnamefont {Adams}}, \bibinfo {author}
  {\bibfnamefont {B.}~\bibnamefont {Amadon}}, \bibinfo {author} {\bibfnamefont
  {T.}~\bibnamefont {Applencourt}}, \bibinfo {author} {\bibfnamefont
  {C.}~\bibnamefont {Audouzec}}, \bibinfo {author} {\bibfnamefont {J.-M.}\
  \bibnamefont {Beuken}}, \bibinfo {author} {\bibfnamefont {J.}~\bibnamefont
  {Bieder}}, \bibinfo {author} {\bibfnamefont {A.}~\bibnamefont {Bokhanchuk}},
  \bibinfo {author} {\bibfnamefont {E.}~\bibnamefont {Bousquet}}, \bibinfo
  {author} {\bibfnamefont {B.}~\bibnamefont {F.}}, \bibinfo {author}
  {\bibfnamefont {D.}~\bibnamefont {Caliste}}, \bibinfo {author} {\bibfnamefont
  {M.}~\bibnamefont {Côté}}, \bibinfo {author} {\bibfnamefont
  {F.}~\bibnamefont {Dahm}}, \bibinfo {author} {\bibfnamefont {F.}~\bibnamefont
  {Da~Pieve}}, \bibinfo {author} {\bibfnamefont {M.}~\bibnamefont {Delaveau}},
  \bibinfo {author} {\bibfnamefont {M.}~\bibnamefont {Di~Gennaro}}, \bibinfo
  {author} {\bibfnamefont {B.}~\bibnamefont {Dorado}}, \bibinfo {author}
  {\bibfnamefont {C.}~\bibnamefont {Espejo}}, \bibinfo {author} {\bibfnamefont
  {G.}~\bibnamefont {Geneste}}, \bibinfo {author} {\bibfnamefont
  {L.}~\bibnamefont {Genovese}}, \bibinfo {author} {\bibfnamefont
  {A.}~\bibnamefont {Gerossier}}, \bibinfo {author} {\bibfnamefont
  {M.}~\bibnamefont {Giantomassi}}, \bibinfo {author} {\bibfnamefont
  {Y.}~\bibnamefont {Gillet}}, \bibinfo {author} {\bibfnamefont
  {D.}~\bibnamefont {Hamann}}, \bibinfo {author} {\bibfnamefont
  {L.}~\bibnamefont {He}}, \bibinfo {author} {\bibfnamefont {G.}~\bibnamefont
  {Jomard}}, \bibinfo {author} {\bibfnamefont {J.}~\bibnamefont
  {Laflamme~Janssen}}, \bibinfo {author} {\bibfnamefont {S.}~\bibnamefont
  {Le~Roux}}, \bibinfo {author} {\bibfnamefont {A.}~\bibnamefont {Levitt}},
  \bibinfo {author} {\bibfnamefont {A.}~\bibnamefont {Lherbier}}, \bibinfo
  {author} {\bibfnamefont {F.}~\bibnamefont {Liu}}, \bibinfo {author}
  {\bibfnamefont {I.}~\bibnamefont {Lukačević}}, \bibinfo {author}
  {\bibfnamefont {A.}~\bibnamefont {Martin}}, \bibinfo {author} {\bibfnamefont
  {C.}~\bibnamefont {Martins}}, \bibinfo {author} {\bibfnamefont
  {M.}~\bibnamefont {Oliveira}}, \bibinfo {author} {\bibfnamefont
  {S.}~\bibnamefont {Poncé}}, \bibinfo {author} {\bibfnamefont
  {Y.}~\bibnamefont {Pouillon}}, \bibinfo {author} {\bibfnamefont
  {T.}~\bibnamefont {Rangel}}, \bibinfo {author} {\bibfnamefont {G.-M.}\
  \bibnamefont {Rignanese}}, \bibinfo {author} {\bibfnamefont {A.}~\bibnamefont
  {Romero}}, \bibinfo {author} {\bibfnamefont {B.}~\bibnamefont {Rousseau}},
  \bibinfo {author} {\bibfnamefont {O.}~\bibnamefont {Rubel}}, \bibinfo
  {author} {\bibfnamefont {A.}~\bibnamefont {Shukri}}, \bibinfo {author}
  {\bibfnamefont {S.}~\bibnamefont {M}}, \bibinfo {author} {\bibfnamefont
  {M.}~\bibnamefont {Torrent}}, \bibinfo {author} {\bibfnamefont
  {M.}~\bibnamefont {Van~Setten}}, \bibinfo {author} {\bibfnamefont
  {B.}~\bibnamefont {Van~Troeye}}, \bibinfo {author} {\bibfnamefont
  {M.}~\bibnamefont {Verstraete}}, \bibinfo {author} {\bibfnamefont
  {D.}~\bibnamefont {Waroquiers}}, \bibinfo {author} {\bibfnamefont
  {J.}~\bibnamefont {Wiktor}}, \bibinfo {author} {\bibfnamefont
  {B.}~\bibnamefont {Xu}}, \bibinfo {author} {\bibfnamefont {A.}~\bibnamefont
  {Zhou}}, \ and\ \bibinfo {author} {\bibfnamefont {J.}~\bibnamefont
  {Zwanziger}},\ }\href@noop {} {\bibfield  {journal} {\bibinfo  {journal}
  {Comput. Phys. Commun.}\ }\textbf {\bibinfo {volume} {205}},\ \bibinfo
  {pages} {106–310} (\bibinfo {year} {2016})}\BibitemShut {NoStop}%
\bibitem [{dp()}]{dp}%
  \BibitemOpen
  \href@noop {} {}\bibinfo {howpublished}
  {\url{https://etsf.polytechnique.fr/Software/Ab_Initio}}\BibitemShut
  {NoStop}%
\end{thebibliography}

\end{document}